\documentclass[
reprint,
amsmath,amssymb,
aps,
prl
]{revtex4-1}
\usepackage{graphicx}

\usepackage{xr}

\usepackage{dcolumn}
\usepackage{bm}
\usepackage{xcolor}
\usepackage{relsize}
\usepackage{enumerate}
\usepackage{xr}
\newcommand*{\addFileDependency}[1]{
\typeout{(#1)}
\@addtofilelist{#1}
\IfFileExists{#1}{}{\typeout{No file #1.}}
}\makeatother
\newcommand{\rom}[1]{\uppercase\expandafter{\romannumeral #1\relax}}

\makeatletter
\newcommand*\bigcdot{\mathpalette\bigcdot@{.5}}
\newcommand*\bigcdot@[2]{\mathbin{\vcenter{\hbox{\scalebox{#2}{$\m@th#1\bullet$}}}}}
\makeatother

\makeatletter
\def\maketitle{
\@author@finish
\title@column\titleblock@produce
\suppressfloats[t]}
\makeatother

\begin{document}

\preprint{APS/123-QED}

\title{Nonlinear Dynamics of Coupled-Resonator Kerr-Combs}

\author{Swarnava Sanyal$^{1}$, Yoshitomo Okawachi$^{1,\dagger}$, Yun Zhao$^{1}$, Bok Young Kim$^{1}$, Karl J. McNulty$^{2}$, 
Michal Lipson$^{1,2}$ and Alexander L. Gaeta$^{1,2}$
}
\email{Corresponding author: a.gaeta@columbia.edu }
\affiliation{%
$^{1}$Department of Applied Physics and Applied Mathematics, Columbia University, New York, New York 10027, USA\\
$^{2}$
Department of Electrical Engineering, Columbia University, New York, New York 10027, USA\\
$^{\dagger}$
Now at Xscape Photonics Inc.
}%
\begin{abstract}
The nonlinear interaction of a microresonator pumped by a laser has revealed complex dynamics including soliton formation and chaos. Initial studies of coupled-resonator systems reveal even more complicated dynamics that can lead to deterministic modelocking and efficient comb generation. Here we perform theoretical analysis and experiments that provide insight into the dynamical behavior of coupled-resonator systems operating in the normal group-velocity-dispersion regime. Our stability analysis and simulations reveal that the strong mode-coupling regime, which gives rise to spectrally-broad comb states, can lead to an instability mechanism in the auxiliary resonator that destabilizes the comb state and prevents mode-locking. We find that this instability can be suppressed by introducing loss in the auxiliary resonator. We investigate the stability of both single- and multi-pulse solutions and verify our theoretical predictions by performing experiments in a silicon-nitride platform. Our results provide an understanding for accessing broad, efficient, relatively flat high-power mode-locked combs for numerous applications in spectroscopy, time-frequency metrology, and data communications.
\end{abstract}
\maketitle
Recent advancements in fabrication of high-quality-factor (high-$Q$) microresonators have enabled the generation of optical Kerr frequency combs driven by a continuous-wave (CW) pump laser. The dynamics of these systems can be accurately modelled using the Lugiato-Lefever Equation (LLE) \cite{Lugiato_PhysRevLett.58.2209,Coen:13_LLE,Chembo_2013_PhysRevA.87.053852,HanssonWabnitz_2016}. This has drawn significant interest from wide-ranging areas of research in science and technology, including spectroscopy \cite{DCS_Vahala_2016,MJ_2018silicon,Dutt_2018}, optical clocks \cite{OpticalClocks_Papp:14}, low-noise microwave generation \cite{lucas2020ultralow,tetsumoto2021optically,zhao2024all,kudelin2024photonic,sun2024integrated}, 
ranging \cite{ranging_Science_2018_doi:10.1126/science.aao3924,ranging_Vahala_2018_doi:10.1126/science.aao1968}, and data communications \cite{Levy_2012_6218758,pfeifle2014coherent,marin2017microresonator, corcoran2020ultra}. Much of the existing research has focused on modelocking via excitation of dissipative Kerr solitons (DKS) in the anomalous group-velocity dispersion (GVD) regime \cite{kippenberg_doi:10.1126/science.aan8083,pasquazi2018micro,gaeta2019photonic}. It has been shown that modelocked combs can also occur in the normal-GVD regime, where non-solitonic solutions to the LLE were found through numerical simulations \cite{OEwaves_Matsko:12,OEwaves_Liang:14,Godey_PhysRevA.89.063814,Parra-Rivas_2016_DarkSolitonPhysRevA.93.063839}. The time-domain profile of such combs is formed via interlocking of switching waves connecting the upper and lower homogenous steady-state solutions of the system \textcolor{black}{\cite{Parra-Rivas_SwitchingWave:16, Parra-Rivas_dark_bright_PhysRevA.95.053863, Vahala_switching_2022}}. Furthermore, under the influence of mode-coupling near the pump wavelength, the system can support comb states referred to as platicons \cite{Lobanov:15,lobanov_2017_beta3_PlaticonDynamics}. Since parametric oscillation, which initiates the comb generation, occurs on the thermally unstable red-detuned branch in the normal-GVD regime \cite{Carmon:04, Xue2016Review}, comb initiation requires the use of pump modulation \cite{lobanov_modulatedPump_2015generation,Lobanov_PhysRevA.100.013807}, laser injection locking \cite{lihachev2022platicon_selfinjectionlocking} or mode-coupling to locally shift the cavity resonance that is pumped \cite{xue2015mode_NatPhot,Jang:16,ScottPapp_yu2022continuum}. 

While conventional anomalous-GVD soliton combs typically suffer from low pump-to-comb conversion efficiencies \cite{herr2014temporal,Jang:21} and an exponentially decaying spectral envelope [Fig \ref{fig1}a (i)], normal-GVD Kerr-combs generated with a coupled-resonator geometry [Fig \ref{fig1}a (ii)], typically exhibit high conversion efficiencies, high optical power per line, and relatively flat spectral profiles \cite{xue_LPR_HighCE_2017microresonator,Kim:19, helgason_dissipative_2021}, which make them ideal for applications such as data communications \cite{Okawachi:23,rizzo2023massively}. 
The interaction between the modes in the main and auxiliary resonators results in mode splitting induced by periodic avoided mode crossings (AMCs) that modifies the dispersion for the single resonator \cite{Zeng:14,Gentry:14} \textcolor{black}{close to the pump}, allowing for the phase-matching of initial modulation-instability (MI) sidebands. 
\textcolor{black}{The system thus supports optical parametric oscillator (OPO) states and single- and multi-pulse comb states, where the later can be accessed deterministically, contrary to the soliton combs.} 
While numerical simulations based on modified coupled-resonator Ikeda Map \cite{Bok_Synchro_doi:10.1126/sciadv.abi4362,helgason_dissipative_2021} offer excellent agreement with experiments, there has been little intuition on what determines the bandwidth and stability of such combs and the conditions for modelocking. \par
\begin{figure*}
\includegraphics{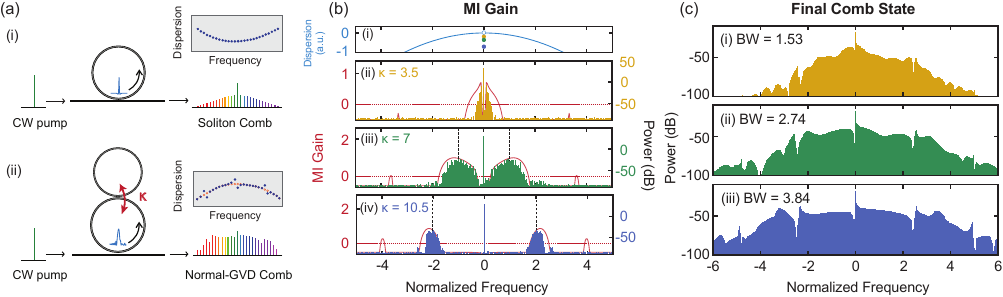}
\caption{\label{fig1} (a) Schematic illustrations of (i) anomalous-GVD soliton comb generated from a single microresonator and (ii) normal-GVD comb generated from a coupled-resonator. The inset shows the corresponding dispersion operators, where for (ii), the periodic AMCs modify the dispersion of the main resonator (originally represented by red circles), allowing for MI to occur. (b) \textcolor{black}{(i) Integrated dispersion profile showing mode-coupling induced shift at the pump and (ii)-(iv)} MI gain spectra (red traces) for three different normalized mode-interaction strengths $\kappa$, with the oscillation threshold denoted by the dotted line (red), which matches the initial growth of sidebands in the output spectrum from numerical simulations (yellow, green, and blue traces). The dotted line (black) shows the location of gain peak obtained using a simplified single-ring model with pump mode-coupling (c) Numerically simulated output comb states corresponding to the three $\kappa$ values \textcolor{black}{[(ii)-(iv) in (b)]}, where the main and auxiliary detunings are optimized for the largest comb bandwidth. The inset shows the corresponding comb bandwidth (in normalized units).
}
\end{figure*}
In this Letter, we perform an in-depth theoretical analysis of the nonlinear dynamics and comb generation in the coupled-resonator system. We illustrate a pathway to increasing the optical bandwidth of a normal-GVD Kerr-comb by introducing a strong AMC. We find that in such a regime, the nonlinearity of the auxiliary cavity can adversely impact the comb stability. We perform a small-signal-gain analysis, which yields insight into the mechanism responsible for destabilizing the comb state in the strong mode-interaction regime and show how this instability can be alleviated by introducing loss into the auxiliary microresonator. We extend our analysis to investigate the stability of the multi-pulse comb states and verify these results via full numerical simulations and experimental demonstrations. Our analysis offers guidance to producing spectrally broad and stable modelocked combs with high conversion efficiencies and relatively flat spectral profiles. 

The normalized coupled-mean field equations used to analyze our coupled-resonator system \cite{xue2019super,ZhuZhao_PhysRevA.105.013524} in the normal-GVD regime are, 
\begin{align}
    \frac{\partial E_1}{\partial t} &= i \kappa E_2 -\left[1 + i \Delta_1 +i \frac{\partial^2}{\partial \tau^2} - i|E_1|^2 \right] E_1 + i S \label{coupled_LLE_norm_1_main}\,,\\
    \frac{\partial E_2}{\partial t} &= i \kappa E_1 - \left[ \alpha + i \Delta_2 +i r \frac{\partial^2}{\partial \tau^2} - i r |E_2|^2 + d \frac{\partial}{\partial \tau} \right] E_2\,, \label{coupled_LLE_norm_2_main}
\end{align}
where $t$, $\tau$ are the slow and fast time variables, normalized to main resonator, $E_{1,2} (t,\tau)$ and $\Delta_{1,2}$ are the normalized intracavity fields and cold-cavity detunings for the main and auxiliary resonators, respectively, $\kappa$ is the strength of the mode-interaction, and $S$ is the pump field amplitude. The auxiliary loss may be different from the main and is given by $\alpha$, $r$=$L_2/L_1$ is a scaling parameter, and $d$ corresponds to a normalized group-velocity mismatch parameter due to the free-spectral range (FSR) difference between the two resonators. The relationship between the normalized and real parameters is given in the Supplementary.

We analyze the dispersive effect of the mode-coupling on the comb bandwidth and find that a stronger mode-interaction strength results in a broader comb spectrum for a given GVD value. This can be understood by looking at the MI behavior in our coupled-resonator system. We perform a linear stability analysis about the continuous-wave steady-state solution of the system \cite{ZhuZhao_PhysRevA.105.013524}. Figure \ref{fig1}b [(ii)-(iv) red traces] shows the MI gain spectra for different $\mathrm{\kappa}$ values, where the degeneracy point for the AMC is located close to the pump. The primary (secondary) MI gain lobes are broader and stronger (narrower and weaker) and correspond to the gain experienced by modes primarily in the main (auxiliary) resonator. The initial intracavity dynamics, as modeled using the coupled Ikeda map, is thus dominated by the growth of modes in the main resonator [yellow, green and blue traces in Fig \ref{fig1}b (ii)-(iv)], which are above the oscillation threshold (gain = 0). \textcolor{black}{Figure \ref{fig1} b(i) shows the integrated dispersion of the single resonator in normal-GVD regime (blue) and the pump resonance shift for each $\mathrm{\kappa}$}. We find that the MI gain frequency moves away from the pump frequency for larger $\kappa$, due to the stronger mode-coupling-induced dispersion \cite{Jang:16}. 

We derive a simple analytical expression for the case of a single normal-GVD ring where the effect of pump mode-coupling is incorporated phenomenologically, by modifying the cold-cavity detuning (Supplementary Section \rom{2}). The location of the gain peak using this simplified model is found to increase with $\kappa$ [vertical dotted line in Fig \ref{fig1}b]. This principle also governs the bandwidth of the normal-GVD comb states. Figure \ref{fig1}c shows the final output comb spectra corresponding to each $\kappa$ value in Fig. \ref{fig1}b \textcolor{black}{[(ii)-(iv)]}, where the main and auxiliary detunings are optimized to yield the broadest comb state for the respective splitting strength \textcolor{black}{(Supplementary Section \rom{9})}. We find that adjusting the splitting-strength results in much greater changes to the comb spectral profile, than changing the detuning values which has been reported for both soliton \cite{Lucas2017_PhysRevA.95.043822} and normal-GVD combs \cite{ScottPapp_yu2022continuum,rebolledo2023platicon}. To better quantify the bandwidth of a normal-GVD comb, we propose a bandwidth definition that considers the entire spectral profile. \textcolor{black}{3-dB and 10-dB bandwidths are useful to characterize soliton-like combs where the spectrum monotonically decays from the center. However, for many frequency combs, including normal-GVD Kerr combs, the center of spectral envelope is ambiguous. We adopt a more general definition of the comb bandwidth as the normalized first-order correlation function of the spectral amplitude.} (Supplementary Section \rom{3}). Using this definition, the comb bandwidth is found to increase with $\kappa$ (see inset of Fig. \ref{fig1}c), while the pump-to-comb conversion efficiency remains roughly constant ($\sim$ 25$\%$). It is important to note that the interaction strength cannot be increased indefinitely for given device parameters since it leads to an increase in the threshold power for comb generation due to the excess mode-coupling loss experienced by the pump \textcolor{black}{in the main resonator \cite{Zhao:21}.}

We investigate the modelocking of combs in the strong mode-interaction regime by performing a linear stability analysis about the pump mode of the comb state in the coupled-resonator system. We focus on specific parameter values similar to the device used in our experiments, but the instability mechanism we observe is universal over a wide range of device parameters. The main and auxiliary FSRs are 100 GHz and 101.42 GHz, respectively, which corresponds to an AMC period of 53 nm. The mode-splitting strength is 6 GHz, which corresponds to a large ring-ring power coupling coefficient of 3.5$\mathrm{\%}$ (compared to below 0.1 $\mathrm{\%}$ for previous demonstrations \cite{xue_LPR_2015normal,Kim:19,helgason_dissipative_2021}). While a strong resonator coupling of nearly 40$\mathrm{\%}$ was implemented in \cite{Vahala_Ji:23}, the FSR values of the two rings were closely matched, which results in a broadband mode interaction that completely reshapes the global dispersion and the system exhibits different dynamics. 
\begin{figure}
\includegraphics{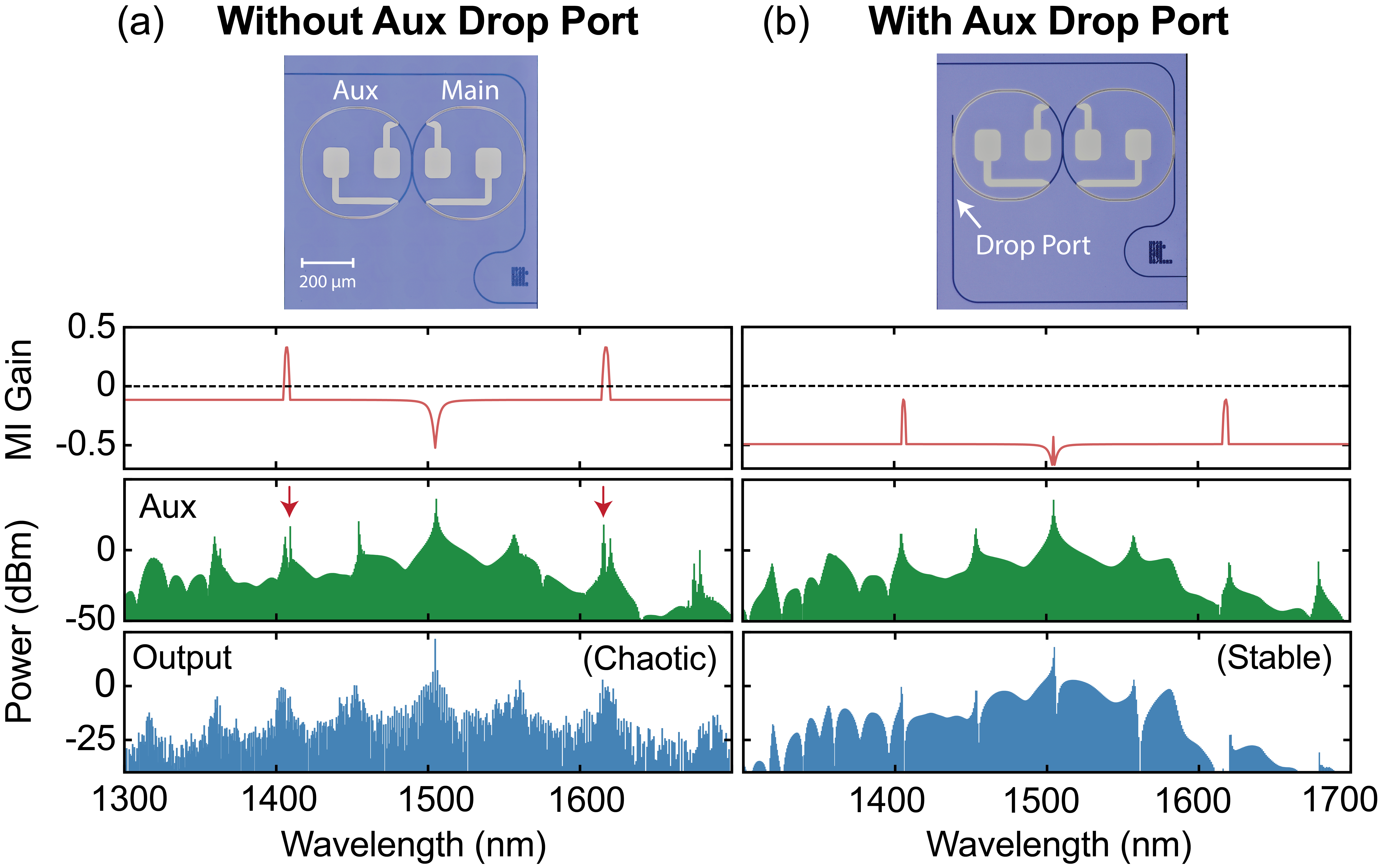}
\caption{\label{fig2} MI Gain spectra (red) and the simulated auxiliary (green) and output (blue) spectra are shown for a coupled-resonator device (a) with and (b) without auxiliary drop port, along with the corresponding microscope images. Above threshold (gain = 0) gain in the auxiliary resonator results in emergence of parasitic MI sidebands in the auxiliary spectrum (indicated by red arrows) in the absence of a drop port, which results in a chaotic comb at the output. A stable modelocked comb state is observed with the drop port, since the excess loss brings the auxiliary MI gain below threshold.}
\end{figure}

Since accessing the comb state requires positioning the AMC close to the pump mode, it results in an excessive power accumulation in the auxiliary resonator for a strong mode-splitting strength. Since the auxiliary resonator has a net lower loss compared to the main resonator, our theory predicts the occurrence of strong MI gain sidebands above the oscillation threshold ($\mathrm{Re\{\lambda\} > 0}$), where the associated eigenvectors imply a growth of modes in the auxiliary resonator [Fig \ref{fig2}a (top)]. This parasitic MI gain mechanism has a deleterious effect on the comb stability, which we verify by performing numerical simulations using a coupled Ikeda map. We access the unstable comb state by reducing the auxiliary nonlinearity by one-fourth which suppresses the instability mechanism and then perform MI gain analysis about the pump mode of this state. The four-wave mixing contributions due to other modes can be phenemenologically added to our model (see Supplementary Section \rom{5}). On reverting the auxiliary nonlinearity to initial value, the previously obtained steady-state auxiliary spectrum, which is multi-peaked due to periodic AMCs, now exhibits emergence of parametric sidebands at the locations predicted by the MI gain spectrum [Fig. \ref{fig2}a (middle)]. These MI sidebands destabilize the comb in the auxiliary and main resonators resulting in a chaotic multi-peaked output spectrum [Fig. \ref{fig2}a (bottom)]. We note that the primary source for this instability is the excess pump power in the auxiliary resonator due to strong mode-interaction. Implementing a drop port on the auxiliary resonator increases its net loss and reduces the power in the pump mode, which lowers the maximum MI gain below the oscillation threshold [Fig. \ref{fig2}b (top)]. Stable modelocking is now observed with the auxiliary and output comb spectra shown in Fig. \ref{fig2}b - middle and bottom, respectively. For the comb states shown in Fig. \ref{fig1}c, an auxiliary drop port was also implemented to inhibit this excess buildup of power.
\begin{figure}
\includegraphics{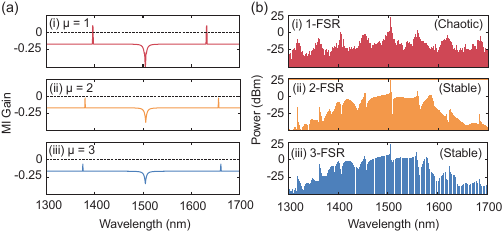}
\caption{\label{fig3} (a) MI gain spectrum and the corresponding (b) simulated output spectrum, without the auxiliary drop port, for (i) 1-FSR, (ii) 2-FSR and (iii) 3-FSR, with inset in (a) showing the location of the AMC from the pump. The 1-FSR state, with the AMC closest to the pump, is destabilized due to the above-threshold auxiliary MI gain, while stable modelocking is observed for higher multi-FSR states.}
\end{figure}
\begin{figure*}[ht]
\includegraphics{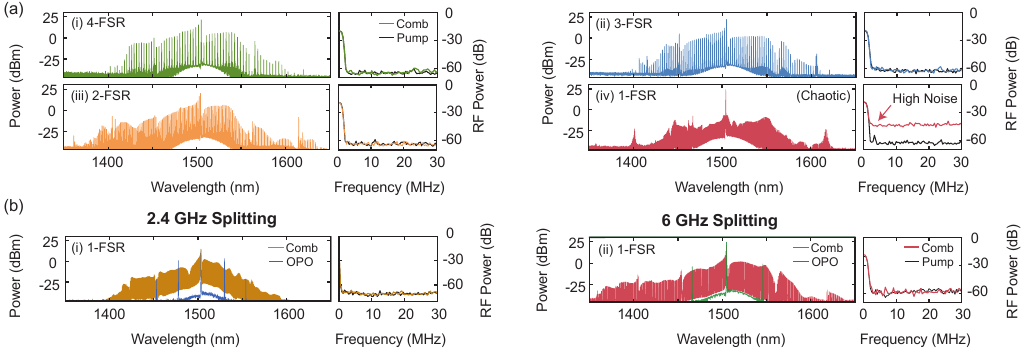}
\caption{\label{fig4} (a) Experimentally measured comb spectra generated using a 100-GHz SiN coupled-resonator device without the auxiliary drop port for a strong mode-splitting strength of 6 GHz, with the inset showing the corresponding noise spectrum measured on an electronic spectrum analyzer. 
Comb spacings of (i) 4-FSR, (ii) 3-FSR, (iii) 2-FSR and (iv) 1-FSR are accessed. While the multi-FSR combs are stable, the 1-FSR state has a high RF-noise, as expected from numerical simulations (Fig. \ref{fig3}). (b) A stable low-noise 1-FSR state is accessible in the device with the auxiliary drop port (ii) for the same splitting strength of 6 GHz. The widest OPO state generated when the AMC is close to the pump is also shown. (i) A device with a weaker mode-splitting strength of 2.4 GHz results in a smaller OPO separation and a narrower 1-FSR comb state.}
\end{figure*}

Analogous to the multi-soliton solutions in the anomalous-GVD regime, it is possible to excite multiple pulses in the normal-GVD regime that correspond to combs with multi-FSR spacings. The number of pulses is dictated by the frequency separation between the AMC position and the pumped cavity resonance, where a higher (lower) number of pulses is obtained when the AMC location is further away from (closer to) the pump \cite{xue_LPR_2015normal,Sanyal:23}. This implies that the comb states corresponding to a higher pulse number have a lower pump power being coupled into the auxiliary resonator. Thus, the multiple pulse states have a lower MI gain peak and are more resistant to the parasitic instabilities, even in the absence of the drop port. We verify this by extending our small-signal-gain analysis to investigate the stability of the multi-FSR comb states. Figure \ref{fig3}a show the gain spectra for 1-FSR, 2-FSR and 3-FSR states, with the legend indicating the number of modes away the AMC is from the pump mode. The corresponding output comb spectra, obtained from numerical simulations, are shown in Fig. \ref{fig3}b. We find that the auxiliary MI gain peak grows weaker with higher pulse number and is below the oscillation threshold for 2-FSR and 3-FSR states, resulting in stable mode-locked combs. The 1-FSR state, with the AMC point closest to the pump, is susceptible to the parasitic instabilities resulting in a chaotic output. Note that at higher mode-splitting strengths or a more overcoupled main resonator, the states with gain peaks just below the oscillation threshold can also become unstable [see Supplementary sections \rom{6}, \rom{7}]. Thus, a careful consideration of the nonlinear dynamics in the coupled-resonator system is essential to access stable modelocked normal-GVD Kerr combs in the strong mode interaction regime.

To verify our theoretical findings, we experimentally compare the normal-GVD Kerr combs generated from two similar silicon nitride (SiN) coupled-resonator devices, with and without the auxiliary drop port [Figs. \ref{fig2}a and \ref{fig2}b show the corresponding microscope images]. The waveguide cross-section is 620 nm $\mathrm{\times}$ 1800 nm, corresponding to a GVD of $\mathrm{35.5\, ps^2/km}$ at the pump wavelength of 1505 nm in numerical simulation. We implement a racetrack design where the long interaction length allows for operation in the strongly overcoupled and large mode-interaction regime. The coupling gaps between the bus waveguide and the main resonator and between the main and auxiliary resonators are 375 nm. For the drop-port device, the gap between the auxiliary resonator and the drop waveguide is 425 nm. The main and auxiliary resonance locations can be \textcolor{black}{adjusted} using the integrated resistive heaters, which allows for tuning the AMC position relative to the fixed pump wavelength and access single and multi-FSR normal-GVD comb states.

We investigate comb generation in the SiN device without the auxilizary drop port using 330 mW of pump power in the bus waveguide. The main resonator has loaded  and intrinsic $Q$ of $\mathrm{3.6 \times 10^5}$ and $\mathrm{5.3 \times 10^6}$, respectively, and the mode-splitting between the two microresonators is measured to be 6.8 GHz at 1545 nm. The auxiliary heater power is initially set such that the AMC location is approximately 4 FSR away, to the red side of the pump. The main heater is then tuned to blueshift the main resonance towards the pump wavelength, resulting in generation of primary MI sidebands followed by a transition to a mode-locked 4-FSR comb state [Fig. \ref{fig4}a (i)]. The auxiliary heater power is then increased, which redshifts the auxiliary resonance. Since the auxiliary resonator has a larger FSR compared to that of the main, the AMC location is blueshifted and approaches the pump wavelength from the red side with increasing auxiliary heater power. This tuning approach allows for deterministic and repeatable access to the lower multi-FSR modelocked states \cite{Sanyal:23}. Spectra (ii) and (iii) in Fig. \ref{fig4}a show stable 3-FSR and 2-FSR states, respectively, with the corresponding RF-noise spectra. As the auxiliary heater power is further increased, we observe a transition to a chaotic 1-FSR state [Fig. \ref{fig4}a (iv)], where the inset shows a high RF noise compared to the pump. The spectral profile is multi-peaked at the locations of periodic AMCs, similar to the simulated output spectrum from Fig. \ref{fig2}a (bottom). The experiment was repeated at lower pump powers without observing a stable 1-FSR state. We conclude that in the strong mode-interaction regime, the 1-FSR comb state is prone to parasitic instabilities without the drop port.  

As predicted by our theoretical analysis, we find experimentally that introduction of a drop port on the auxiliary resonator minimizes the parasitic instability and allows for a stable 1-FSR state. For this device, the main resonator has loaded and intrinsic $Q$ of $\mathrm{4.2 \times 10^5}$ and $\mathrm{5.5 \times 10^6}$, respectively, with a mode-splitting strength of 6.59 GHz measured at 1535 nm. A low-noise 1-FSR comb state is observed [Fig. \ref{fig4}b (ii)] with 62 lines above 0.5 mW and a pump-to-comb conversion efficiency of 33.5$\mathrm{\%}$ for a pump power of 370 mW. To further illustrate that a stronger AMC allows access to broader and spectrally flatter comb states, we compare comb generation in a similar device but with a weaker mode-splitting strength of 2.4 GHz. The blue and green traces in Fig. \ref{fig4}b (i) and (ii) show the widest optical parametric oscillator (OPO) state generated in the two respective devices when the AMC is brought close to the pump. \textcolor{black}{A stronger mode-coupling results in a larger separation between the MI gain peaks [Fig. \ref{fig1}b] and thus, a wider OPO state}. Due to a much lower mode-interaction-induced loss at the pump mode, a lower pump power of 178 mW was used to access the 1-FSR state [Fig. \ref{fig4}b (i)] in the weaker splitting device with a similar conversion efficiency of 32.9 $\mathrm{\%}$. The comb spectra shown in Figs. \ref{fig4}b (i) and (ii) are vastly different, even though the two devices are nearly identical except for the ring-ring coupling strength, which agrees with our numerical simulation [Fig. \ref{fig1}c]. 

In summary, we theoretically and experimentally study the stability of modelocked normal-GVD Kerr-combs in a coupled-resonator system. Our analysis predicts an occurrence of a parasitic instability in the strong mode-coupling regime, due to excessive power accumulation in the auxiliary resonator. The single-FSR comb states, with the AMC location closer to the pump, are found to be more susceptible to this instability mechanism, resulting in a chaotic output. A drop port that reduces the pump power in the auxiliary ring is crucial for stabilizing the high-power comb states. Operating in this strong mode-interaction regime allows us to achieve comb-line power equalization over a broad spectral region while maintaining high power per comb line, making the normal-GVD Kerr-combs ideal for applications such as data communication and spectroscopy. 
\begin{acknowledgments}
The authors acknowledge support from the Advanced Research Projects Agency - Energy (DE-AR0000843), the Air Force Office of Scientific Research (FA9550-15-1-0303), and the Defense Advanced Research Projects Agency (HR0011-19-2-0014 and HR0011-18-3-0002). This work was performed, in part, at the Cornell NanoScale Facility, a member of the National Nanotechnology Infrastructure Network, which was supported by the National Science Foundation.
\end{acknowledgments}

\newpage
\title{Supplementary Material - Nonlinear Dynamics of Coupled-Resonator Kerr-Combs}
\maketitle
\onecolumngrid
\setcounter{equation}{0}
\setcounter{figure}{0}
\setcounter{table}{0}
\renewcommand{\theequation}{S\arabic{equation}}
\renewcommand{\thefigure}{S\arabic{figure}}
\renewcommand{\bibnumfmt}[1]{[S#1]}
\renewcommand{\citenumfont}[1]{S#1}

\section{\rom{1}.\hspace{0.3cm} Coupled-Ring Modulation Instability Analysis} \label{section_Coupled_ring_MI}
\begin{figure}
\includegraphics{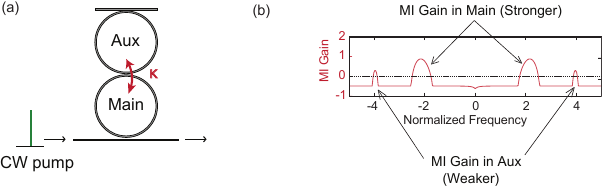}
\caption{\label{CR_gain} (a) Coupled resonator system with an auxiliary drop port driven by a CW pump. (b) MI Gain spectrum for $\kappa = 10.5$ as shown in Fig 1(b) of the main text. The eigenvectors associated with the stronger (weaker) gain peaks overlaps with the modes in main (auxiliary) resonator. }
\end{figure}

Modulation Instability (MI) in a single ring driven by a continuous-wave (CW) pump has been extensively analyzed. \cite{haelterman1992dissipative_sm,haelterman1992additive_sm}. In this section, we analyze the MI in a coupled resonator system shown in Fig. \ref{CR_gain}a. Our system is modeled using the coupled-mean field equations which can be written as
\begin{align}
    T_{R1} \frac{\partial E_1(t,\tau)}{\partial t} &= \left[ -\alpha_1 - i \delta_1 +i L_1 \sum_{k\geq2}\frac{\beta_k}{k!}\left(i\frac{\partial}{\partial \tau}\right)^k + i \gamma_1 L_1 |E_1|^2 \right] E_1 + i \sqrt{\theta_2}E_2 + i \sqrt{\theta_1} E_{in}     \label{Coupled_LLE_real_1}\\
    T_{R1} \frac{\partial E_2(t,\tau)}{\partial t} &= \left[ -\alpha_2 - i \delta_2 +i L_2 \sum_{k\geq2}\frac{\beta_k}{k!}\left(i\frac{\partial}{\partial \tau}\right)^k + i \gamma_2 L_2 |E_2|^2 -\delta L_1 \frac{\partial}{\partial\tau}\right] E_2 + i \sqrt{\theta_2}E_1 
    \label{Coupled_LLE_real_2}
\end{align}

where $E_1(t,\tau)$ and $E_2(t,\tau)$ are the intracavity field envelopes for the main and auxiliary resonators respectively (their absolute value squared is normalized to power), $t$ is the slow time variable that tracks the field evolution and $\tau$ is the fast time variable spanning the roundtrip time of the main resonator.  The coefficients $\alpha_1 (\alpha_2)$, $\delta_1 (\delta_2)$, $L_1 (L_2)$, $T_{R1} (T_{R2})$ correspond to total cavity roundtrip loss, cold-cavity detuning, cavity length and roundtrip time for main (auxiliary) resonator. $\beta_k$ denote the $k^{th}$ order dispersion coefficient and the two resonators are assumed to have different nonlinearity coefficients $\gamma_1$ and $\gamma_2$ respectively (The need for this will become clear in Section \ref{section_MI_gain_evaluation}). $E_{in}$ is the input CW pump field driving the main resonator, $\theta_1$ is the input power coupling coefficient and $\theta_2$ is the power coupling coefficient between the two resonators. The auxiliary resonator has a drop port with a power coupling coefficient $\theta_3$. The total cavity losses are thus written as $\alpha_1 = (\alpha_{i1} L_1+\theta_1)/2$  and $\alpha_2 = (\alpha_{i2}L_2+\theta_3)/2$, where $\alpha_{i1}$ ($\alpha_{i1}$) is the intrinsic loss per unit length for main (auxiliary) resonator. The coupled equations are written in the frame of reference of the main resonator and since there is a slight free-spectral-range (FSR) mismatch between main and auxiliary resonator ($\mathrm{FSR_1}$ and $\mathrm{FSR_2}$ denotes the individual FSRs respectively), the field equation for the auxiliary resonator has a drift term with the coefficient \textcolor{black}{$\delta = 1/L_1(1/\mathrm{FSR_2} - 1/\mathrm{FSR_1})$}. 

For MI gain analysis, it is easier to work in normalized units \cite{Coen:13_UniversalScalingLaws_sm, ZhuZhao_PhysRevA.105.013524_sm}. We define dimensionless slow and fast time variables and the field amplitude as follows:
\begin{equation}
    t \rightarrow \alpha_1 \frac{t}{T_{R1}}, \quad \tau \rightarrow \tau \sqrt{\frac{2\alpha_1}{|\beta_2|L_1}}, \quad  E_{1,2} \rightarrow E_{1,2} \sqrt{\frac{\gamma L_1}{\alpha_1}}, \quad
\end{equation}

The normalized mean-field equations can thus be written as

\begin{align}
    \frac{\partial E_1}{\partial t} &= \left[ -1 - i \Delta_1 +i  \sum_{k\geq2} \eta^{(k)}_1\left(i\frac{\partial}{\partial \tau}\right)^k + i |E_1|^2 \right] E_1 + i \kappa E_2 + i S \label{coupled_LLE_norm_1}\\
    \frac{\partial E_2}{\partial t} &= \left[ -\alpha - i \Delta_2 +i  \sum_{k\geq2} \eta^{(k)}_2\left(i\frac{\partial}{\partial \tau}\right)^k + i\, r \,\Gamma\,  |E_2|^2 - d \frac{\partial}{\partial \tau} \right] E_2 + i \kappa E_1 \label{coupled_LLE_norm_2}
\end{align}
where we define following normalized parameters
\begin{align}
    \Delta_1 &= \frac{\delta_1}{\alpha_1},\quad \Delta_2 = \frac{\delta_2}{\alpha_1},\quad \alpha = \frac{\alpha_2}{\alpha_1},\quad \kappa = \frac{\sqrt{\theta_2}}{\alpha_1},\quad d = \delta \sqrt{\frac{2 L_1}{|\beta_2| \alpha_1}}, \quad \Gamma = \frac{\gamma_2}{\gamma_1}, \\
    S &= E_{in}\sqrt{\frac{\gamma L_1 \theta_1} {\alpha^3_1}},\quad r = \frac{L_2}{L_1},\quad \eta^{(k)}_1 = \frac{\beta_k L_1}{\alpha_1 k!} \left( \frac{2\alpha_1}{|\beta_2|L_1}\right)^{k/2},\quad \eta^{(k)}_2 = r \cdot \eta^{(k)}_1
\end{align}

The steady-state homogenous solutions $E_{1ss}$ and $E_{2ss}$ of the system can be found by setting the temporal derivatives equal to zero and are given by solutions to the following coupled algebraic equations
\begin{align}
    [-1 - i\Delta_1 + i |E_{1ss}|^2)]E_{1ss} + i\kappa E_{2ss} + i S &= 0 \label{cw_ss_1}\\
    [-\alpha - i\Delta_2 + i \,r\,\Gamma\,|E_{2ss}|^2)]E_{2ss} + i\kappa E_{1ss} &= 0 \label{cw_ss_2}
\end{align}

We can then perform linear stability analysis of the intracavity fields $E_1(t,\tau)$ and $E_2(t,\tau)$ about the steady-state solutions by writing

\begin{align}
    E_1(t,\tau) &= E_{1ss} + a_s(t)e^{-i\Omega\tau} + a_i(t)e^{i\Omega\tau}\\
    E_2(t,\tau) &= E_{2ss} + b_s(t)e^{-i\Omega\tau} + b_i(t)e^{i\Omega\tau}
\end{align}
where $a_s (b_s)$ and $a_i (b_i)$ are the small-signal amplitudes for the signal and idler modes of main (auxiliary) resonator at the modulation frequency $\Omega$. This ansatz is inserted into equations \ref{coupled_LLE_norm_1},\ref{coupled_LLE_norm_2}. Only even orders of dispersion survive and we retain terms upto $\mathrm{4^{th}}$ order in dispersion. Ignoring the terms which are of order $\mathrm{\geq 2}$ in the small-signal amplitudes and retaining , we obtain the following rate equations:
\begin{align}
    \frac{\partial a_{s}}{\partial t} &=  \left( -1 -i \Delta_1 + i \eta^{(2)}_1 \Omega^2 + i \eta^{(4)}_1 \Omega^4 + i\, 2 |E_{1ss}|^2\right) a_{s} + i E^2_{1ss}a^*_{i} + i \kappa b_{s} \label{eqn_main_signal}\\
    \frac{\partial a_{i}}{\partial t} &=  \left( -1 -i \Delta_1 + i \eta^{(2)}_1 \Omega^2 + i \eta^{(4)}_1 \Omega^4 + i\, 2 |E_{1ss}|^2\right) a_{i} + i E^2_{1ss}a^*_{s} + i \kappa b_{i}\label{eqn_main_idler}\\
    \frac{\partial b_{s}}{\partial t} &=  \left( -\alpha -i \Delta_2 + i \eta^{(2)}_1 \Omega^2 + i \eta^{(4)}_1 \Omega^4 + i\, 2\,r\,\Gamma\,|E_{2ss}|^2\right) b_{s} + i\,r\,\Gamma\, E^2_{2ss}b^*_{i} + i \kappa a_{s} + i\,\Omega d b_s\label{eqn_aux_signal}\\
    \frac{\partial b_{i}}{\partial t} &=  \left( -\alpha -i \Delta_2 + i \eta^{(2)}_1 \Omega^2 + i \eta^{(4)}_1 \Omega^4 + i\, 2\,r\,\Gamma\,|E_{2ss}|^2\right) b_{i} + i\,r\,\Gamma\, E^2_{2ss}b^*_{s} + i \kappa a_{i} - i\,\Omega d b_i\label{eqn_aux_idler}
\end{align}

The above equations can be written in matrix form as follows:
\begin{equation}
    \begin{pmatrix}
        \dot{a}_s\\
        \dot{a}^*_i\\
        \dot{b}_s\\
        \dot{b}^*_i
    \end{pmatrix}
    =
    \underbrace{\begin{pmatrix}
        -1 + i\,\Delta k_1/2 & i E_{1ss}^2 & i \kappa & 0\\
        -i E_{1ss}^{*2} & -1 - i\,\Delta k_1/2 & 0 & -i \kappa\\
        i \kappa & 0 & -\alpha + id\Omega + i\,\Delta k_2/2 & ir\Gamma E_{2ss}^2\\
        0 & -i \kappa & -ir\Gamma E_{2ss}^{*2} & -\alpha + id\Omega - i\,\Delta k_2/2
    \end{pmatrix}}_{\mathlarger{\mathrm{M}}}
    \begin{pmatrix}
        {a}_s\\
        {a}^*_i\\
        {b}_s\\
        {b}^*_i
    \end{pmatrix}  
    \label{eqn_matrix_MI}
\end{equation}

where $\Delta k_1 = 4 |E_{1ss}|^2 - 2 \Delta_1 + 2\eta^{(2)}_1\Omega^2 + 2\eta^{(4)}_1\Omega^4$ and $\Delta k_2 = 4\,r\,\Gamma\, |E_{2ss}|^2 - 2 \Delta_2 + 2\eta^{(2)}_2\Omega^2 + 2\eta^{(4)}_2\Omega^4$.

The largest real part of eigenvalues for the above $4\times4$ matrix $\mathrm{M}$ ,$\mathrm{Re(\lambda})$, gives the MI gain. The CW steady-state solution of the system is first obtained by solving the coupled equations (\ref{cw_ss_1},\ref{cw_ss_2}). 
In our case, the detuning of main resonator $\Delta_1$ is scanned from blue to red, so the system thus stays on the upper branch of the steady-state solution.  We keep the detuning of the auxiliary resonator $\Delta_2$, that determines the location of the avoided mode crossing (AMC), close to the pump which ensures the pump mode resonance experiences a sufficient mode-coupling induced shift to initiate the MI process. For a given set of system parameters, the MI gain spectrum is obtained by plotting the eigenvalue with the largest real part for each modulation frequency $\Omega$. As opposed to a single ring, the MI gain spectrum for a coupled ring system can exhibit two set of peaks symmetric to the pump, where the associated eigenvector implies which resonator experiences the growth of signal and idler modes. If the eigenvector lies in the subspace spanned by $\{\left( 1 ,0, 0 ,0 \right) , \left( 0 ,1, 0 ,0 \right) \}$, the modes in main resonator experience the associated MI gain, whereas if it lies in the subspace spanned by $\{\left( 0 ,0, 1 ,0 \right) , \left( 0 ,0, 0 ,1 \right) \}$, the auxiliary resonator modes would start oscillating.
\\
Fig 1b of the main text shows that the peaks in the MI gain spectrum move away from the pump for stronger normalized mode interaction strength $\kappa$. The normalized system parameters chosen are as follows: scaling-factor $r = L_2/L_1$ = 0.975, the dispersion-coefficients $\eta_{1,2}^{(2)}$ = (1.0, 0.975), $\eta_{1,2}^{(4)}$ = (0,0), the loss for the auxiliary ring $\alpha$ = 0.493, pump power $S$ = 1.837. 
The periodicity of the AMC, determined by the drift parameter $d$, is chosen to be sufficiently large ($d$ = -142.703) to ensure the intracavity dynamics near the pump frequency remains unaffected by the mode-interaction at the next period. 

The main and auxiliary detunings were chosen so that the AMC location is close to the pump: (i) $\kappa$ = 3.48: $(\Delta_1, \Delta_2)$ = (2.64, 11.18), (ii) $\kappa$ = 6.97: $(\Delta_1, \Delta_2)$ = (5.30, 13.85) and (iii) $\kappa$ = 10.46: $(\Delta_1, \Delta_2)$ = (8.43, 16.97). The large value of $|d|$ ensures that that the intracavity dynamics around the pump is only affected by the AMC closest to it and not at the next period. Figure \ref{CR_gain}b shows the MI gain spectrum for case (iii). As mentioned above, by examining the associated eigenvector, we find that the stronger (weaker) MI gain lobes correspond to growth of modes in the main (auxiliary) resonator. The coupled-ring MI gain spectra has certain additional features which are different from the single ring case. First, the background for the single-ring MI is at $-1$ which is the normalized loss term. For the coupled-ring system, the loss term for main and auxuiliary resonators is different ($-1$ and $-\alpha$ respectively with $|\alpha|<1$ for an asymmetric drop port). Since we plot the largest real part of the eigenvalues for the MI gain spectrum, the coupled-ring MI background is at $-\alpha$. Secondly, due to the AMC induced loss near the pump mode, the background also becomes more negative close to the pump.  

A broader final comb state is obtained for a stronger mode-interaction strength $\kappa$. Fig 1c of the main text shows the final comb states for different $\kappa$ values, where the main and auxiliary detunings were optimized to acheive the broadest comb state for each case: (i) $\kappa$ = 3.48: $(\Delta_1, \Delta_2)$ = (4.56, 7.27), (ii) $\kappa$ = 6.97: $(\Delta_1, \Delta_2)$ = (7.73, 10.39) and (iii) $\kappa$ = 10.46: $(\Delta_1, \Delta_2)$ = (12.29, 12.12). 

\section{\rom{2}.\hspace{0.3cm} Analytical expression for pump mode-coupling assisted MI}
\label{section_gain_maxima_eqn}
The MI gain spectrum obtained by linearizing about the homogenous solutions of the coupled mean-field equations agrees well with our numerical simulations, as shown in Fig. 1b of the main text. In this section, we derive a simple analytical expression showing the dependence of the frequency location of the MI gain peak ($\Omega_{max}$) on the mode-coupling strength $\kappa$. This is done by  reducing the coupled-ring system to an equivalent single-ring system in the normal-GVD regime, where the effect of pump mode-coupling is included phenomenologically as a shift in the cold-cavity detuning just for the pump mode \cite{Jang:16_sm}. This modifies the homogenous steady-state solution $\overline{E}_{1ss}$ of the single-ring system, now given by
\begin{align}
    [-1 - i\Delta_s + i |\overline{E}_{1ss}|^2)]\overline{E}_{1ss} + i S &= 0 \label{cw_ss_jae_field}
\end{align}

where $\Delta_s = \Delta_1 - \kappa_0$ is the shifted pump detuning and $\Delta_1$ is the unperturbed detuning. $\kappa_0$ corresponds to the AMC induced shift at the pump in this reduced single-ring model and is related to the ring detunings $\Delta_1$ and $\Delta_2$ and mode-coupling strength $\kappa$ for the full coupled-ring model as follows
\begin{align}
    \kappa_0 &= \sqrt{\frac{(\Delta_2 - \Delta_1)^2}{4} + \kappa^2} - \frac{(\Delta_2 - \Delta_1)}{2} \label{effective_AMC_shift}
\end{align}

The intracavity power $\overline{Y} = |\overline{E}_{1ss}|^2$, from eqn. \ref{cw_ss_jae_field}, is given by the solution to the well-known cubic polynomial \cite{haelterman1992additive_sm,haelterman1992dissipative_sm} with the shifted-detuning $\Delta_s$ and pump power $X = |S|^2$
\begin{align}
    \overline{Y}^3 - 2 \Delta_s \overline{Y}^2 + (\Delta_s^2 + 1)\overline{Y} &= X \label{cw_ss_jae_power}
\end{align}

The rate equations for the signal and idler small-signal amplitudes are given by

\begin{equation}
    \begin{pmatrix}
        \dot{a}_s\\
        \dot{a}^*_i
    \end{pmatrix}
    =
    \begin{pmatrix}
        -1 + i\,\Delta k/2 & i \overline{E}_{1ss}^2\\
        -i \overline{E}_{1ss}^{*2} & -1 - i\,\Delta k/2 
    \end{pmatrix}
    \begin{pmatrix}
        {a}_s\\
        {a}^*_i
    \end{pmatrix}  
    \label{eqn_matrix_MI_jae}
\end{equation}

where $\Delta k = 4 |\overline{Y}|^2 - 2 \Delta_1 + 2\Omega^2 $ is the phase-mismatch paramter, that depends explicitly on the unshifted detuning $\Delta_1$ and the effect of mode-coupling is included implicitly through the modified steady-state solution $\overline{Y}$. 

\begin{figure}
\includegraphics{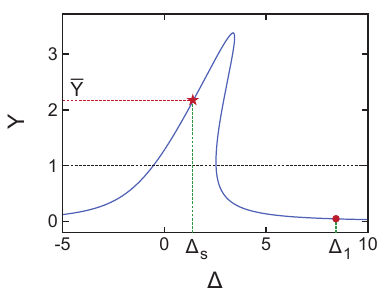}
\caption{\label{cw_ss_mode_coupling} Homogeneous steady-state solution for a single-ring system, where the mode-coupling shifts the cold-cavity detuning $\Delta_1$ to $\Delta_s$, causing the system to cross the oscillation threshold ($\mathrm{Y = 1}$)}
\end{figure}

Figure \ref{cw_ss_mode_coupling} shows how the AMC induced shift affects the steady-state solution of the system. In absence of mode-coupling, the solution $Y(\Delta_1)$ is below the MI gain threshold ($Y = 1$). But when a detuning shift of $\kappa_0$ is introduced, the solution crosses the oscillation threshold ($Y(\Delta_s) = \overline{Y} > 1$) allowing for the MI phase-matching conditions to be satisfied. The location of the MI gain peak is then given by perfect phase-matching $\Delta k = 0$ :
\begin{align}
    \Omega_{max} &= \sqrt{(\Delta_1 - 2 \overline{Y})}\\ \label{omega_max}
    &= \sqrt{(\Delta_s + \kappa_0 - 2 \overline{Y})}
\end{align}

Thus, $\Omega_{max}$ increases monotonically with $\kappa_0$, where 
for different $\kappa_0$ values, the system parameters can be adjusted so as to operate at same shifted detuning and intracavity power $(\Delta_s , \overline{Y})$. This is done by adjusting the ring detunings $\Delta_1$ and $\Delta_2$ in the full coupled-ring system such that $(\Delta_2 - \Delta_1)$ and $\Delta_s$ are fixed. This also fixes the AMC location (close to the pump) for different $\kappa_0$ . From Eqn \ref{effective_AMC_shift}, $\kappa_0$ then depends only on the mode-coupling strength $\kappa$. Thus, the MI gain peak would move farther away from the pump for a stronger $\kappa$. 

In Fig. 1b of the main text, we compare the peak MI gain location obtained from eqn \ref{omega_max} (dotted line) with the MI gain spectrum obtained from the full coupled-ring gain analysis. Our reduced single-ring model is able to closely predict the location of MI gain peak. As explained earlier, for different $\kappa$, the ring detunings are chosen such that $(\Delta_2 - \Delta_1)$ and $\Delta_s$ are fixed (see parameters in Section \ref{section_Coupled_ring_MI}). The predicted $\Omega_{max}$ is found to be slightly less than the actual peak location, because the reduced model doesn't include the loss experienced by the pump due to AMC and thus overestimates the intracavity power $(\overline{Y})$, resulting in a smaller value for $\Omega_{max}$ from eqn. \ref{omega_max}. Note that for $\kappa = 3.48$, the gain maxima is found to overlap with the pump mode and isn't shown in the figure.

\section{\rom{3}.\hspace{0.3cm} General definition for Comb Bandwidth}
\label{section_comb_bw_def}
\begin{figure}
\includegraphics{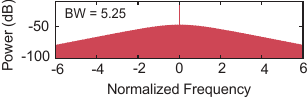}
\caption{\label{bw_def} Soliton comb spectrum with normalized bandwidth calculated using eqn \ref{eqn_bw} given in the inset. }
\end{figure}

In this section, we propose a definition of optical bandwidth for an arbitrary comb shape. The existing definitions of comb bandwidth, such as 3-dB or 10-dB bandwidths, are useful for soliton-like combs with a monotonically decaying spectrum (sech-squared profile). For coupled-resonator-based normal-GVD combs, where the spectrum can have slow modulation far away from the pump owing to the periodic mode interaction, we propose a bandwidth definition $\Delta \Omega$ that considers the entire spectral profile,
\begin{equation}
\Delta \Omega =  \sum_{\mu} |G_\mu|^2 \,\cdot\,\mathrm{FSR}\;\;, 
\label{eqn_bw}
\end{equation}
where
\begin{equation}
G_\mu = \frac{\sum\limits_{\nu\neq0}|E_\nu|\cdot|E_{\nu+\mu}|}{\sum\limits_{\nu\neq0}|E_\nu|^2}\;\;,
\end{equation}
Here $\nu=0$ corresponds to the pump mode index and $G_\mu$ is the normalized first-order correlation function for the spectral amplitude. Using this definition in Figure 1c of the main text, the comb bandwidth (in normalized units) is found to increase with $\kappa$. For comparison, we use this definition to quantify the bandwidth of a soliton comb ($\Delta$ = 4.70 and $S$ = 1.96 \cite{Coen:13_UniversalScalingLaws_sm}) shown in Figure \ref{bw_def}. The 10-dB bandwidth for the soliton is 5.06 (in normalized units), while the calculated bandwidth from eqn. \ref{eqn_bw} is 5.25. Comparing this with Fig 1c of the main text, we note that though the soliton comb has a broader bandwidth, the pump-to-comb conversion efficiency is limited to few percent, limiting it's application for use in data communication and spectroscopy.

\section{\rom{4}.\hspace{0.3cm} Coupled Ikeda Map}
\label{section_ikeda_map}
We use a coupled-Ikeda Map model \cite{Bok_Synchro_doi:10.1126/sciadv.abi4362, helgason_dissipative_2021_sm} to simulate the comb dynamics in our coupled-resonator system. We use a nonlinear schr$\mathrm{\ddot{o}}$dinger equation (NLSE) with distributed loss to simulate the field evolution in the cavities and include the effects of ring-bus and ring-ring coupling and linear phase accumulation due to cavity detunings as boundary conditions.  
\begin{figure}[ht]
\includegraphics{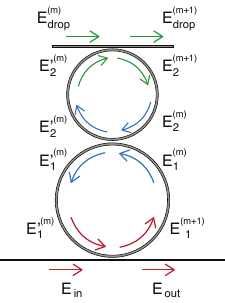}
\caption{\label{SM_IkedaMap} Coupled-ring system modelled using coupled-Ikeda Map with three coupling regions: main ring-bus, main ring-auxiliary ring and auxiliary ring-drop port.}
\end{figure}

Using real units defined in Section \ref{section_Coupled_ring_MI} and intracavity fields at the coupling regions as in Fig. \ref{SM_IkedaMap} , one step of the simulation consists of :

\begin{enumerate}[(i)]

  \item $\mathrm{L_{1}/2}$ propagation in main and auxiliary rings:
  \begin{align}
    \frac{\partial E_1}{\partial z} &= \left[ -\frac{\alpha_{i1}}{2}  +i \sum_{k\geq2}\frac{\beta_k}{k!}\left(i\frac{\partial}{\partial \tau}\right)^k + i \gamma |E_1|^2 \right] E_1     \label{Coupled_NLSE_1}\\
    \frac{\partial E_2}{\partial z} &= \left[ -\frac{\tilde{\alpha}_{i2}}{2} + i \sum_{k\geq2}\frac{\tilde{\beta}_k}{k!}\left(i\frac{\partial}{\partial \tau}\right)^k + i \tilde{\gamma} |E_2|^2 -\delta \frac{\partial}{\partial\tau}\right] E_2 
    \label{Coupled_NLSE_2}
\end{align}
where $(\tilde{\alpha_{i2}},\tilde{\beta_2},\tilde{\gamma_2})$ = $r\, \cdot (\alpha_{i2},\beta_2,\gamma_2)$ are scaled parameters for the auxiliary NLSE with $r$ = $L_2/L_1$.
  \item 
  Ring-Ring coupling and linear phase accumulation for half-roundtrip:
  \begin{equation}
      \begin{bmatrix}
        E'^{(m)}_2\\
        E'^{(m)}_1
    \end{bmatrix}
    =
    \begin{bmatrix}
        \sqrt{1 - \theta_2} & i\,\sqrt{\theta_2} \\
        i\,\sqrt{\theta_2} & \sqrt{1-\theta_2}
    \end{bmatrix}
    \bigcdot
    \begin{bmatrix}
        e^{-i \delta_2/2} & 0 \\
        0 & e^{-i \delta_1/2} 
    \end{bmatrix}
    \begin{bmatrix}
        E^{(m)}_2\\
        E^{(m)}_1
    \end{bmatrix}
  \end{equation}
  \item $\mathrm{L_{1}/2}$ propagation in main and auxiliary rings
 \item 
  Auxiliary ring-drop port coupling and linear phase accumulation for half-roundtrip:
  \begin{equation}
      \begin{bmatrix}
        E^{(m+1)}_{\mathrm{drop}}\\
        E^{(m+1)}_2
    \end{bmatrix}
    =
    \begin{bmatrix}
        \sqrt{1 - \theta_3} & i\,\sqrt{\theta_3} \\
        i\,\sqrt{\theta_3} & \sqrt{1-\theta_3}
    \end{bmatrix}
    \bigcdot
    \begin{bmatrix}
        1 & 0 \\
        0 & e^{-i \delta_2/2} 
    \end{bmatrix}
    \begin{bmatrix}
        E^{(m)}_{\mathrm{drop}}\\
        E'^{(m)}_2
    \end{bmatrix}
  \end{equation}
\item Ring-bus coupling and linear phase accumulation for half-roundtrip :
  \begin{equation}
      \begin{bmatrix}
        E_{out}\\
        E^{(m+1)}_1
    \end{bmatrix}
    =
    \begin{bmatrix}
        \sqrt{1 - \theta_1} & i\,\sqrt{\theta_1} \\
        i\,\sqrt{\theta_1} & \sqrt{1-\theta_1}
    \end{bmatrix}
    \bigcdot
    \begin{bmatrix}
        1 & 0 \\
        0 & e^{-i \delta_1/2} 
    \end{bmatrix}
    \begin{bmatrix}
        E_{in}\\
        E'^{(m)}_1
    \end{bmatrix}
  \end{equation}

 \end{enumerate}

\section{\rom{5}.\hspace{0.3cm} Auxiliary MI Gain in the strong mode interaction regime in absence of the drop port}
\label{section_MI_gain_evaluation}
In this section, we explain how one can investigate the parasitic MI gain mechanism in the auxiliary resonator that leads to comb instability in the strong mode interaction regime. A rigorous stability analysis of comb states in the coupled resonator system would involve linearizing about the steady-state solution of the coupled mean field equations \ref{coupled_LLE_norm_1},\ref{coupled_LLE_norm_2}. However, we claim that since the predominant source of instability in the strong mode interaction regime arises due to excess power in the pump mode of the auxiliary resonator, our approach of linearizing about the pump mode of the comb state to perform MI gain analysis is sufficient to determine the stability of the comb state and we verify our results with numerical simulations. Thus, in the rate equations for signal and idler of the two resonators (eqn. \ref{eqn_matrix_MI}), we can make the following substitution:
\begin{align}
    E_{1ss} \longrightarrow a_0\;\;\;\;,\;\;\;\;
    E_{2ss} \longrightarrow b_0
    \label{assumption_pumpmode}
\end{align}
where $a_0$ and $b_0$ are the pump mode powers in main and auxiliary resonators respectively, obtained from numerical simulation of the system, which will be different from the steady-state solutions due to nonlinear conversion to other comb modes. Note that the effect due to other comb modes can be phenomenologically added in our model by incorporating all the four-wave mixing interactions from different comb modes that can couple the signal and idler of the main and auxiliary resonators. We will show that these contributions only result in a minor change in the MI gain spectrum. The results from our theoretical analysis accurately predict the behaviour of the system and match with numerical simulations. 

In the rate equations for the small-signal amplitudes for the signal (idler) of main resonator (eqns. \ref{eqn_main_signal},\ref{eqn_main_idler}), the term $i2 |E_{1ss}|^2$ corresponds to cross-phase modulation due to the pump mode in the main resonator while the term $i E^2_{1ss}a^*_{i(s)}$ corresponds to nonlinear four-wave mixing process. The same holds true for the equations for auxiliary ring. The assumption in eqn. \ref{assumption_pumpmode} only includes the contribution from the pump mode of the comb state. The effect due to other modes is included as follows: First, we can include the cross-phase modulation from all the other comb modes, where the power in these modes is evaluated from numerical simulations. Thus, we make the following substitutions for the signal (idler) rate equations

\begin{align}
    i\;2 |E_{1ss}|^2 &\longrightarrow i\; 2 \sum_{\mu \neq s(i)} |a_\mu|^2\\
    i\;2 |E_{2ss}|^2 &\longrightarrow i\; 2 \sum_{\mu \neq s(i)} |b_\mu|^2
    \label{assumption_xpm}
\end{align}

Second, we can include other four-wave mixing interactions involving comb modes symmetrically located with respect to the pump with the following substitution:

\begin{align}
    i\;E^2_{1ss}a^*_{i(s)} \longrightarrow i\;\left[\sum_{\substack{0\leq\mu\leq N/2 \\ \mu \neq i(s)}} a_{-\mu}a_{+\mu}\right] a^*_{i(s)}\\
    i\;E^2_{2ss}b^*_{i(s)} \longrightarrow i\;\left[\sum_{\substack{0\leq\mu\leq N/2\\ \mu \neq i(s)}} b_{-\mu}b_{+\mu}\right] b^*_{i(s)}    
    \label{assumption_nlgain}
\end{align}

where $\mu = 0$ corresponds to the pump mode. We will now compare the MI gain spectrum evaluated by considering the effect of just the pump mode and all the comb modes. But before that, there is an additional challenge to evaluating the MI gain spectrum for an unstable steady-state solution of the system and observing the initial emergence of MI sidebands using numerical evolutionary simulations based on the split-step Fourier algorithm. An exact analytical expression for the unstable steady-state comb solution is absent, making it harder to linearize the coupled mean-field equations about this solution and using it as an initial seed in our numerical simulations. 

\begin{figure}[ht]
\includegraphics{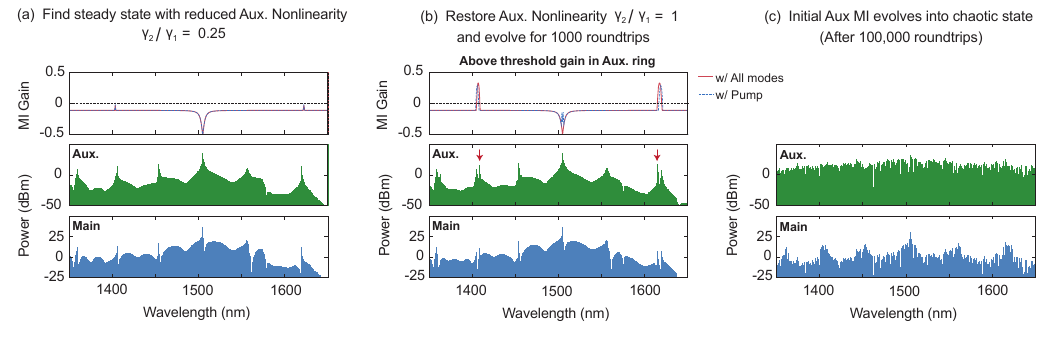}
\caption{\label{unstable_sol_access} Numerical technique to evaluate the MI Gain spectra and observe the initial auxiliary MI sidebands for an unstable comb state (a) The auxiliary nonlinearity is first reduced to one-fourth of that of the main to supress the instability due to strong mode interaction, thereby allowing us to numerically access the otherwise unstable comb solution. MI Gain spectrum is below threshold due to reduced auxiliary nonlinearity. (b) Restore the aux nonlinearity to original value and perform MI gain analysis as explained in the text using the steady state spectrum found in (a). Dotted blue and red traces correspond to MI gain spectra obtained by including the effect of just pump mode and all the comb modes respectively. The auxiliary spectrum exhibits rise of MI sidebands on numerically evolving the system for 1000 roundtrips at the location predicted by the gain analysis, which results in (c) a chaotic comb after 100,000 roundtrips. }
\end{figure}

To overcome this issue, we implement a numerical technique as explained in Figure \ref{unstable_sol_access}. We first numerically access the approximate unstable steady-state solution by reducing the nonlinearity of the auxiliary resonator to one-fourth of that of the main resonator. The steady state main and auxiliary comb solution obtained from numerical simulations is shown in Fig. \ref{unstable_sol_access} (a) blue and green traces respectively. The MI gain analysis can be performed to incorporate just the effect due to pump mode of this comb solution (dotted blue trace) or all the other comb modes (red trace). Due to lower auxiliary nonlinearity, the MI gain is below threshold. Next, the auxiliary nonlinearity is restored to its original value in Fig. \ref{unstable_sol_access}b. Due to this, the MI gain (still evaluated using the comb solution from (a)) crosses the oscillation threshold. The associated eigenvector is found to lie in the subspace spanned by $\{\left( 0 ,0, 1 ,0 \right) , \left( 0 ,0, 0 ,1 \right) \}$, thereby implying that gain is experienced by the auxiliary resonator modes.  On evolving the numerical simulation for 1000 roundtrips, we see emergence of MI sidebands in the auxiliary spectrum at the locations predicted by the gain analysis. The MI gain spectra from red and dotted blue traces only differ slightly, but the former gives a closer match with the numerical simulations. We thus include all the comb modes for our gain analysis in Figs. 2 and 3 of the main text. The initial MI sidebands grow stronger and due to further four-wave mixing interactions with other comb modes result in a chaotic auxiliary comb, thereby destabilizing the comb state in main resonator. Figure \ref{unstable_sol_access}c shows the simulated main and auxiliary spectrum after evolving for 100,000 roundtrips. The former exhibits multiple peaks at the locations of periodic AMCs (every 53 nm).

For this simulation and those shown in Figs. 2 and 3 of the main text, we work with real units for a particular waveguide geometry with the following parameters: $FSR_1$ = 100 GHz, $FSR_2$ = 101.42 GHz, $L_1$ = 1421 $\mu m$, $L_2$ = 1386 $\mu m$, Dispersion: $\beta_2$ = 35.52 $ps^2/km$, $\beta_3$ = -0.27 $ps^3/km$, $\beta_4$ = 0.002375 $ps^4/km$ (at pump wavelength of 1505 nm for a waveguide cross-section of 620 $\times$ 1800 nm), $\gamma_1$ = $\gamma_2$ = 1.229 $\mathrm{(Wm)^{-1}}$, $\alpha_{i1}$ = 2.935 $\mathrm{m^{-1}}$, $\alpha_{i2}$ = 2.8944 $\mathrm{m^{-1}}$, $\theta_1$ = 0.0315, $\theta_2$ = 0.0341, $P_{in}$ = 350 mW. This corresponds to intrinsic $Q_{in\, 1,2}$ = $\omega_0 T_{R1,2}/(2 \alpha_{i1,2} L_{1,2})$ = 3 $\times$ $\mathrm{10^6}$ for main and auxiliary resonators and a main loaded $Q_{l,1}$ of 3.5 $\times$ $\mathrm{10^5}$. For the simulations with the auxiliary drop port, we use $\theta_3$ = 0.0138 which corresponds to an auxiliary loaded $Q_{l,2}$ of 7 $\times$ $\mathrm{10^5}$. We choose an asymmetric drop port in simulations, similar to the device used in the experiments. The main and auxiliary cold-cavity detunings $\delta_{1,2}$ (in radians) are related to the frequency detuning of the individual resonances from the pump laser $\Delta f_{1,2}$ as $\Delta f_{1,2} = \delta_{1,2} \,\mathrm{FSR}_{1,2} / (2 \pi)$. For comb states shown in Fig. \ref{unstable_sol_access} and Fig. 2 of the main text, the frequency detunings are $\Delta f_{1,2}$ = 3.23 , 3.45 GHz. 

The detuning values for multi-FSR comb states shown in Fig. 3 of the main text are: $\Delta f_{1,2}$ = 2.3 , 6 GHz (3-FSR), $\Delta f_{1,2}$ = 2.5 , 5.6 GHz (2-FSR), $\Delta f_{1,2}$ = 3.1 , 4.1 GHz (1-FSR). For the chosen FSR difference between the main and the auxiliary resonators, this corresponds to the AMC located approximately 3-FSR, 2-FSR and 1-FSR away from the pump mode, respectively. For the stable multi-FSR states, MI gain analysis was performed about the pump mode of the steady-state comb obtained from numerical simulation. But for the unstable 1-FSR state, the same approach of lowering the auxiliary nonlinearity, as described above, was used to obtain the steady-state solution for the gain analysis.
As explained in Section \ref{section_Coupled_ring_MI}, the background of the coupled-ring MI gain becomes more negative close to the pump mode. A similar trend is also noted in Fig. 3a of the main text. As the AMC location relative is brought closer to the pump, it induces a stronger mode-coupling loss resulting in a more negative MI gain background close to the pump.

\section{\rom{6}.\hspace{0.3cm} Parasitic Auxiliary Instability for Lower Multi-FSR states}
\label{section-sensitive dependence on detunings}
\begin{figure}[ht]
\includegraphics{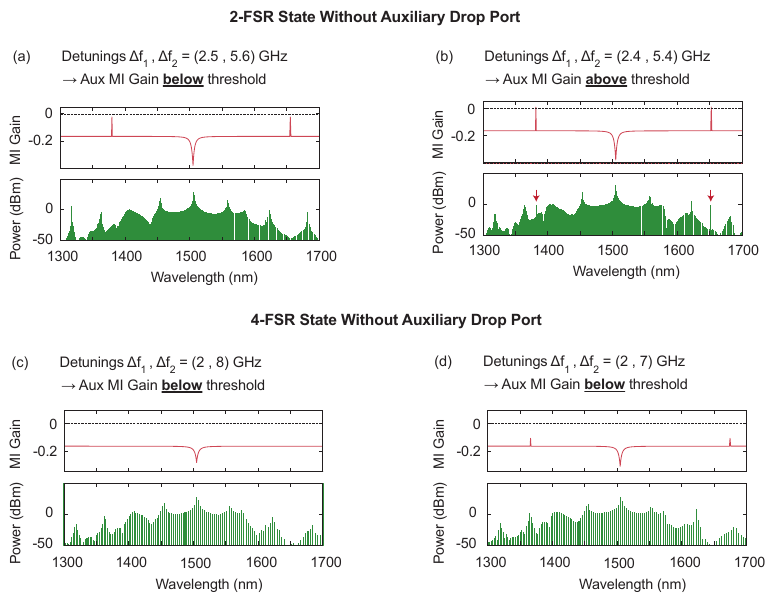}
\caption{\label{sensitive_dependence_2fsr_4fsr} MI Gain spectra (red) and auxiliary spectra (green) for 2-FSR and 4-FSR states in absence of the drop port for two different main and auxiliary detuning values. In (a), the MI gain is below threshold resulting in stable modelocked state. But slight change in detunings, brings the MI gain for the 2-FSR state above oscillation threshold resulting in emergence of parasitic sidebands in the auxiliary spectrum (red arrows). (c) and (d) show two detuning conditions for a 4-FSR state that correspond to the lowest and highest MI gain peak value, respectively, as the detunings are varied. The 4-FSR state remains stable over the whole existence range.}
\end{figure}
In Fig. 3a of the main text, we showed that the MI gain peak becomes weaker with higher multi-FSR or multi-pulse states. For the chosen system parameters and in absence of the auxiliary drop port, the 1-FSR state is always rendered unstable. However, a stable 2-FSR state is found to exist, as shown in Fig. 3 (ii), with detunings of 2.5 GHz and 5.6 GHz, respectively. Figure \ref{sensitive_dependence_2fsr_4fsr} (a) shows the below-threshold MI Gain spectrum (red trace) and the simulated auxiliary spectrum (green trace) for this stable 2-FSR state. However, the MI gain peak is very close to the oscillation threshold and slight change in the detuning values can result in an above threshold gain. This is illustrated in Fig. \ref{sensitive_dependence_2fsr_4fsr} (b) where the MI gain spectrum is shown for slightly different main and auxiliary detunings. Since the gain is positive but small, the system has to be evolved for roughly 20,000 roundtrips to observe the rise of the parasitic MI sidebands in the auxiliary spectrum at the location predicted from our gain analysis. The system eventually evolves into a chaotic comb state.

For comparison, we also show the MI gain and the auxiliary ring spectra for a 4-FSR comb state in Fig. \ref{sensitive_dependence_2fsr_4fsr} (c) and (d) for two different main and auxiliary detuning conditions that correspond to minimum and maximum MI gain peak value for a 4-FSR state, respectively. The MI gain for the 4-FSR state remains far below threshold even when the detunings are varied. Thus, the stability of lower multi-FSR comb states with MI gain close to the oscillation threshold is highly sensitive to the the detuning values, which can make it potentially harder to access such states experimentally. Thus, a careful study of nonlinear dynamics in the coupled-resonator system is essential to check which of the multi-FSR states could be observed in experiments.
\section{\rom{7}.\hspace{0.3cm} Parasitic Auxiliary MI due to heavy overcoupling}
\label{section Aux MI due to overcoupling}
In this section, we show that even for moderate mode-interaction strengths, the parasitic auxiliary instability can occur when the main resonator is heavily overcoupled. This regime of operation is usually implemented to access comb states with high pump-to-comb conversion efficiencies. Here we analyze the instability for a 200-GHz Normal-GVD comb with following parameters: $\mathrm{FSR_1}$ = 200 GHz, $\mathrm{FSR_2}$ = 206 GHz, $L_1$ = 699 $\mu m$, $L_2$ = 678 $\mu m$, $\beta_2$ = 40 $ps^2/km$, $\beta_3$ = -0.27 $ps^3/km$, $\beta_4$ = 0.002375 $ps^4/km$, $P_{in}$ = 400 mW, intrinsic $Q_{in\, 1,2}$ = 3 $\times$ $\mathrm{10^6}$ for main and auxiliary resonators, a main loaded $Q_{l1}$ = 2 $\times$ $\mathrm{10^5}$ and an asymmetric drop port with auxiliary loaded $Q_{l,2}$ = 1.3 $\times$ $\mathrm{10^6}$. We use a mode-splitting strength of 6 GHz ($\theta_2$ = 0.00862), which has been implemented in previous experiments \cite{Kim:19_sm}, but the main resonator was critically coupled or slightly overcoupled. 
\begin{figure}[ht]
\includegraphics{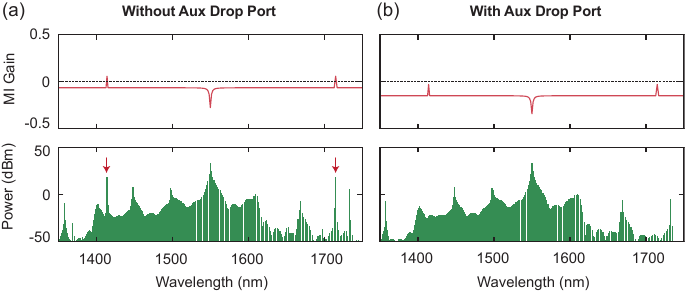}
\caption{\label{high_CE_auxMI} (a) Above threshold auxiliary MI gain (red trace) observed for a heavily overcoupled main resonator with moderate mode-interaction strength ($\theta_2$ = 0.00862) which results in rise of MI sidebands in the auxiliary spectrum (green trace). (b) Drop-port implementation brings the MI gain below threshold resulting in a stable auxiliary spectrum.}
\end{figure}

A 1-FSR comb state is accessed with main and auxiliary frequency detunings $\Delta f_{1,2}$ = 2.5 and 6.2 GHz, respectively. The MI Gain spectra is shown in Figure \ref{high_CE_auxMI}a (red trace) and is found to be above the oscillation threshold. The auxiliary spectrum, shown in Fig. \ref{high_CE_auxMI}a (green trace), exhibits occurence of MI sidebands at the location of MI gain peaks which result in a chaotic comb state on further evolution. The primary cause of instability here is still the excessive pump mode power in the auxiliary resonator, but not due to a strong AMC, but instead due to a heavily overcoupled main resonator. A drop port on the auxiliary resonator brings the MI gain below threshold resulting in a stable auxiliary spectrum [Fig. \ref{high_CE_auxMI}b].

\section{\rom{8}.\hspace{0.3cm} Normal-GVD Pulse Profile}
\label{section_pulse_shape}

\begin{figure}[ht]
\includegraphics{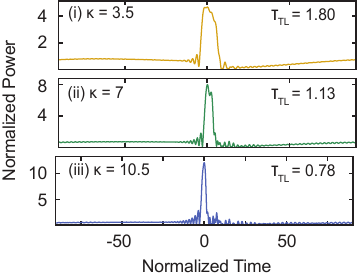}
\caption{\label{pulse_shape} \textcolor{black}{Normalized time-domain pulse profile in the main resonator for the normal-GVD comb states shown in Fig. 1(c) of main text. The transform-limited pulse width $\tau_{TL}$ is shown in the legend.}}
\end{figure}

\textcolor{black}{In a single microresonator in the normal GVD regime, the steady-state pulse solution is a dark pulse state that can be described as interlocking of switching waves connecting the upper and lower homogenous CW solutions \cite{Parra-Rivas_SwitchingWave:16_sm}. The presence of third-order dispersion modifies the switching-wave profiles, allowing the system to support bright modelocked pulses  \cite{Parra-Rivas_dark_bright_PhysRevA.95.053863_sm}. Bright pulse states are also obtained in the presence of mode-coupling in a coupled resonator system, but the upper and lower levels of the pulse shape do not coincide with the CW steady-state solution of the coupled system \cite{helgason_dissipative_2021_sm}. Figure \ref{pulse_shape} shows the normalized time-domain profile in the main resonator for the comb states shown in Fig. 1(c) of the main text for different mode-coupling strengths $\kappa$. Since the normal-GVD pulses are chirped, we also show the transform-limited pulse width in the inset for each pulse state. For a larger $\kappa$ value, the transform-limited pulse width is lower, which is consistent with the broadening of the comb spectral envelope. Thus, shorter pulses with a higher peak-power are obtained in the strong mode-coupling regime.
}

\section{\rom{9}.\hspace{0.3cm} Optimization of Comb Bandwidth for a given Mode-Splitting Strength}
\label{section_bw_optimize}
\begin{figure}[ht]
\includegraphics{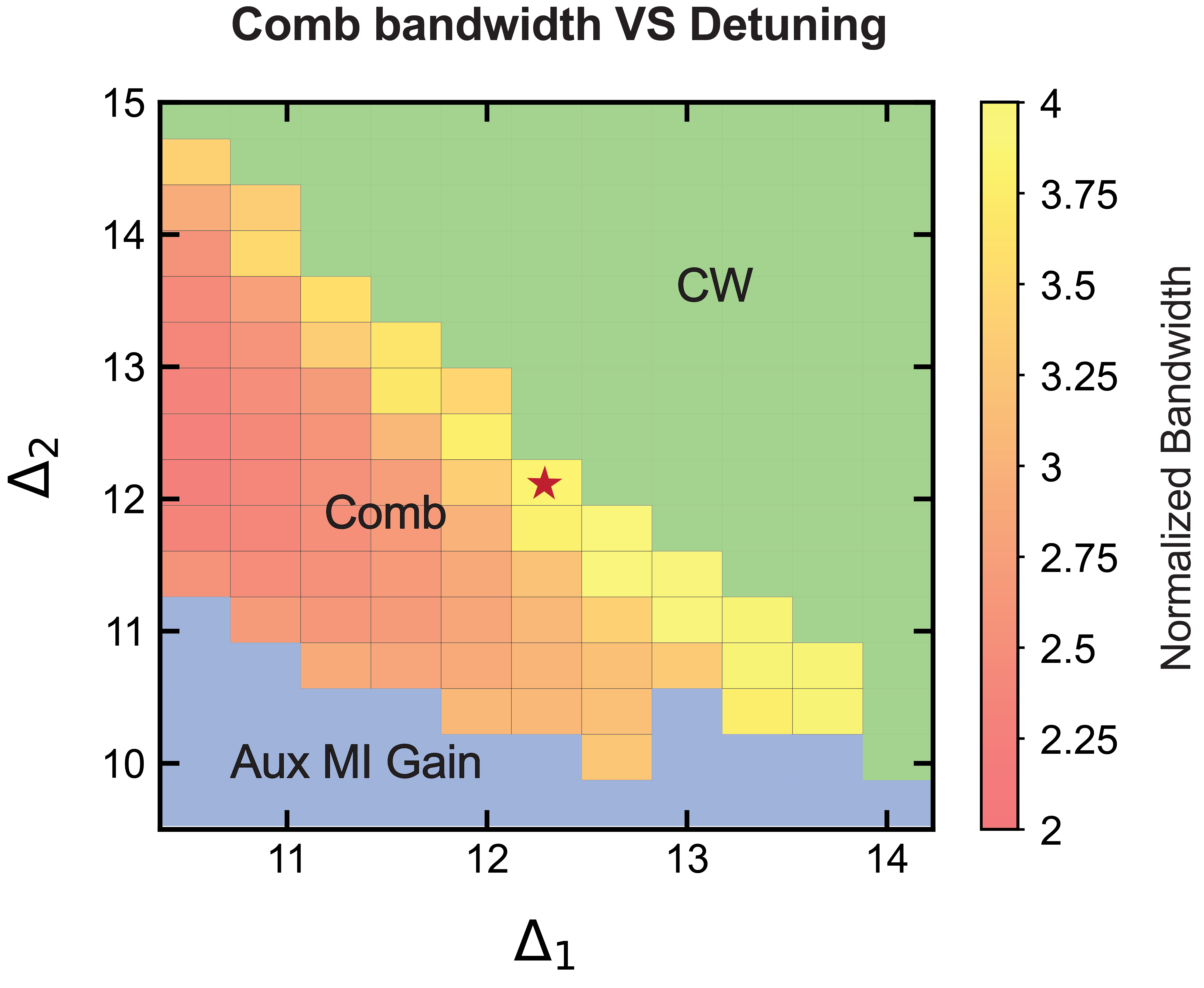}
\caption{\label{bw_optimize} \textcolor{black}{Variation of comb bandwidth as the main and auxiliary detunings are varied. The region in blue and green denote the unstable comb state due to auxiliary MI gain and the CW state, respectively. The point marked in star denotes the comb state shown in Fig. 1c (iii) of the main text.}}
\end{figure}
\textcolor{black}{In this section, we describe our procedure to optimize the main and auxiliary detunings ($\Delta_1, \Delta_2$) to yield the broadest the normal-GVD comb state for a given mode-splitting strength. Using a normal-GVD comb state as a seed, we performed a raster scan where the main and auxiliary detunings were varied such that the mode-crossing location was close to that of the pump mode. The other system parameters were held fixed. For each main and auxiliary detuning value, the final state was recorded after evolving for 10,000 roundtrips, and the corresponding comb bandwidth (calculated using equation \ref{eqn_bw}) was plotted. We show the result of the scan for the case of $\kappa$ = 10.5 in Fig. \ref{bw_optimize}. When $\Delta_1$ and $\Delta_2$ are close to $\kappa$, a near-degenerate splitting occurs and the effective cold-cavity detuning $\Delta_{-} = \frac{\Delta_1 + \Delta_2}{2} - \sqrt{\frac{(\Delta_2-\Delta_1)^2}{4} + \kappa^2}$ (the detuning of the pump from the red-shifted split mode) is nearly zero. Due to excess power in the pump mode of the auxiliary resonator, the comb state is unstable in this regime. The blue shaded region shows the combs susceptible to the auxiliary MI gain. For a given $\Delta_1$, the comb state collapses to the CW solution (shown in green) for a large $\Delta_2$. The comb state shown in Fig. 1c (iii) of the main text with BW = 3.84 is represented by the datapoint marked with a star.
}

%


\end{document}